\documentclass{pasj00}
\draft

\begin{document}
\SetRunningHead{Priest}{Conference Summary}

\def \bv {{\bf v}}
\def \del {{\bf \nabla}}
\def \div {{\bf \del} \cdot}
\def \p {\partial}
\def \. {\cdot}

\title{Hinode 7: 

Conference Summary and Future Suggestions}

\author{Eric \textsc{Priest}} %
\affil{Mathematics Institute, St Andrews University, St Andrews KY16 8QR, UK}
\email{erp@st-andrews.ac.uk}

%


\maketitle

\begin{abstract}
This conclusion to the meeting  attempts to summarise what we have learnt during the conference (mainly from the review talks) about new observations from Hinode and about theories stimulated by them. Suggestions for future study are also offered.
\\
\end{abstract}

\section{Introduction}
We have been treated to an outstanding set of review talks here, and so it is a real pleasure to summarise the main points from them.  What have we learnt from the invited reviews about the big questions in solar physics, and where should we go next? But first an advert and a look back.

For the past 10 years, I have been writing a replacement for the book {\it Solar MHD}.  Three days ago, I finally finished the page proofs, and so it will hopefully appear next spring, published by Cambridge University Press (Priest, 2014). The ``baby" is a completely new rewrite, not just a new edition, so I had to decide on a new name for the new baby. In the end, I came up with {\it Magnetohydrodynamics of the Sun}, so as to indicate that the subject matter is the same as before, but the book is very different.

\begin{figure}
 \begin{center}
  \includegraphics[width=16cm]{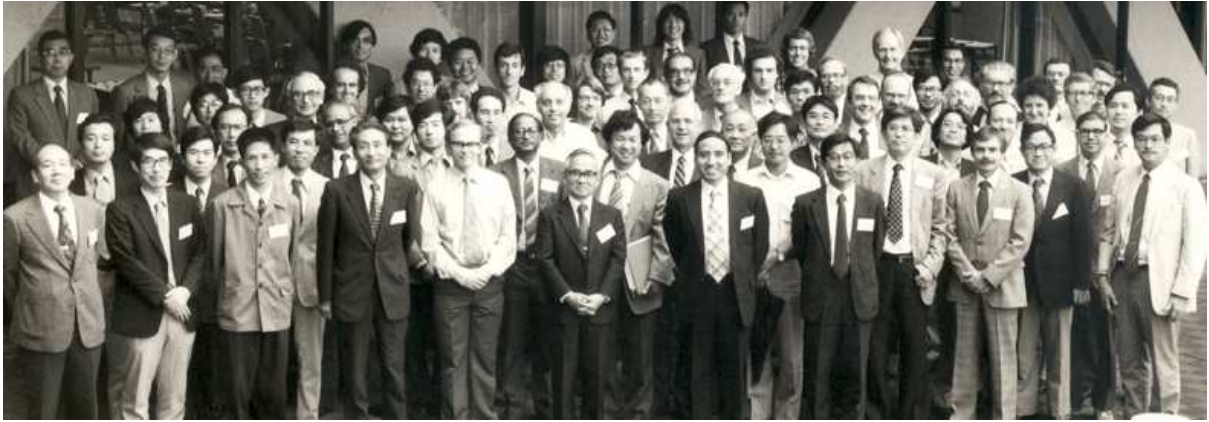} 
 \end{center}
\caption{The Hinotori symposium (Tokyo, 1982).}\label{fig1}
\end{figure}
My first visit to Japan was over 30 years ago in 1982 to the Hinotori symposium in Tokyo, and many key figures in our field can be seen in the conference photograph as rather younger people (Fig. \ref{fig1}). Near the centre of the photograph there is Uchida-san, whom I admired greatly as a highly creative MHD theorist, as well as Watanabe-san and two young graduate students, Sakurai-san (hiding on the back row) and Shibata-san (behind my shoulder), who at the time thought that reconnection has no role in solar flares!  How one's ideas can change over time!  On the left side of the photograph on the second or third row stands Ichimoto-san between Suematsu-san and a serious bespectacled Tsuneta-san. Finally, on the right, you can see Hirayama-san and Hiei-san on the front row, with Doschek, Acton, Svestka and Tandberg-Hanssen a little further back.

\begin{figure}
 \begin{center}
  \includegraphics[width=13cm]{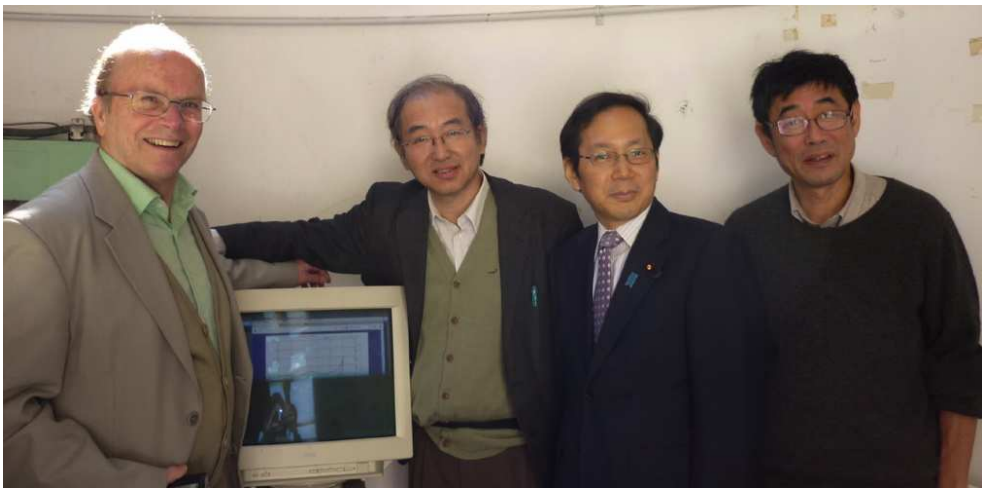} 
 \end{center}
\caption{Shibata-san and Ichimoto-san showing me and a visiting politician an X-class flare at Kwasan observatory.}\label{fig2}
\end{figure}
A few days ago, Ichimoto-san and Shibata-san took me on October 28th to Kwasan observatory, where they and Mai Kamobe kindly laid on a special treat  -- my first X-class flare seen alive in real time. 

Hinode of course continues to make major contributions to fundamental understanding, thanks to teams of selfless scientists working under the brilliant PI's with the instruments and the data, both on SOT, EIS and XRT.
We have over the past few days been treated to an excellent set of talks, so what have we learnt about the big questions and where should we go next?   The individual talks referred to here can be found in these proceedings, and related work that has been published elsewhere is listed at the end of this article.

\section{The Structure of the Convection Zone}

Hideyuki Hotta, a research student with a bright future, described how helioseismology has shown us several features: 

(i) the equator is accelerated due to the transport of angular momentum by the Reynolds stress  $ \langle v_r v_\phi \rangle$;

(ii) the internal velocity in the convection zone is constant on cones, due to a subtle balance including the effects of an entropy gradient and of meridional flow;

(iii) a strong shear layer (the tachocline) is located at the base of the convection zone;

(iv) and a near-surface shear layer, which is not understood at all but may be due to small-scale (granular) convection.

\begin{figure}
 \begin{center}
  \includegraphics[width=7cm]{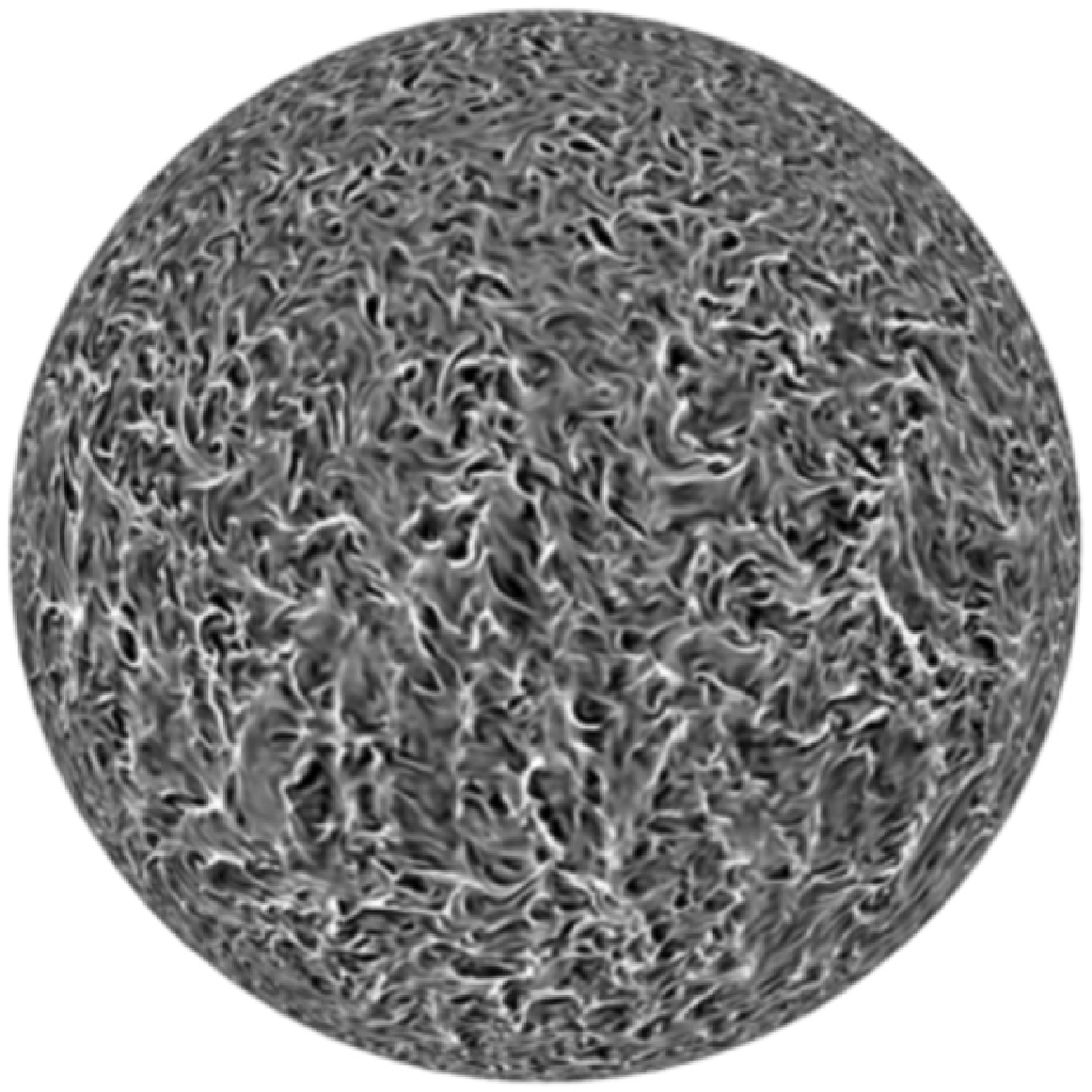}
    \includegraphics[width=7cm]{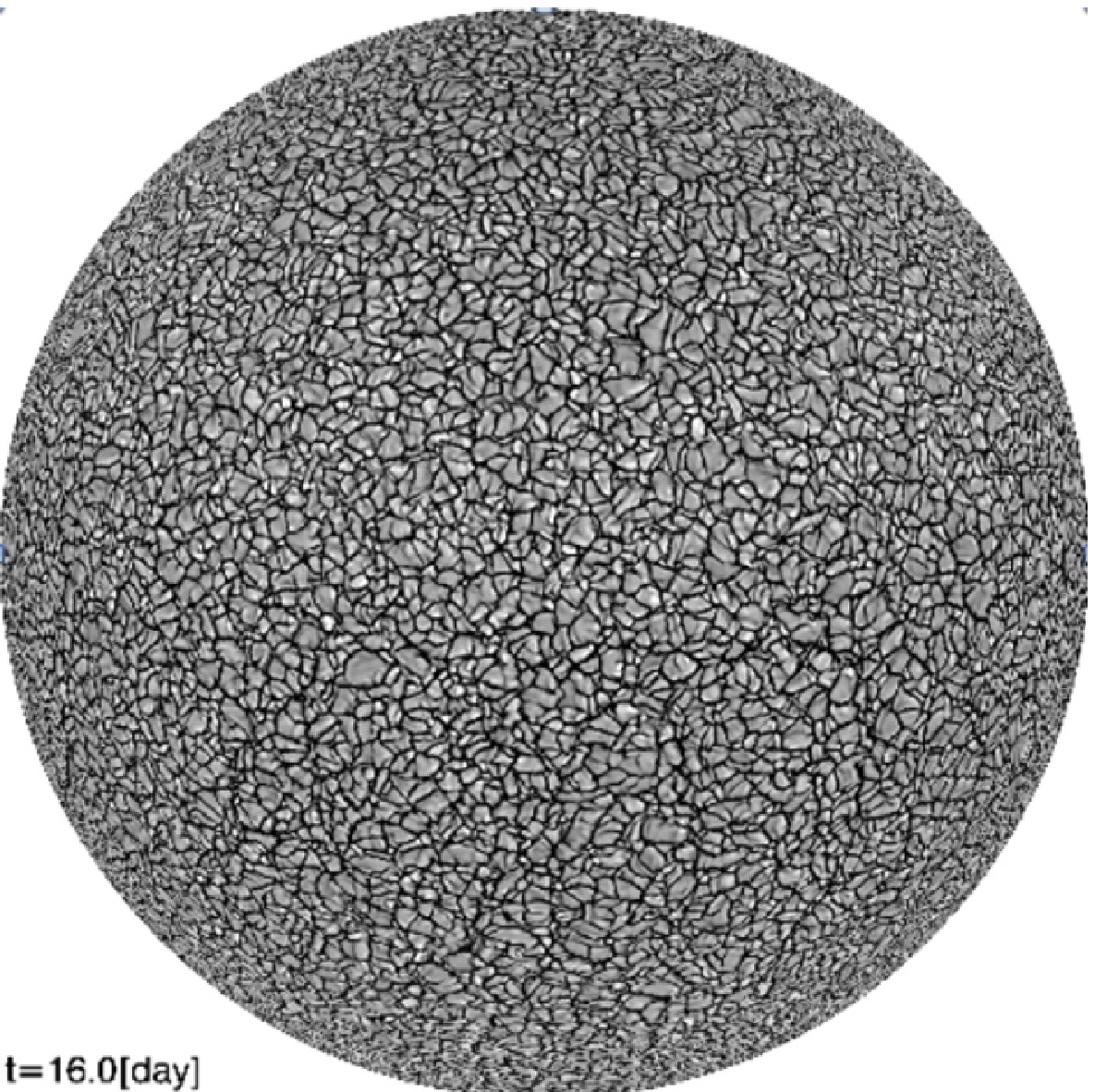}  
 \end{center}
\caption{Global internal convective flow at radii of (a) $r=0.715 R_0$ and (b) $r=0.99 R_0$ in Hotta's model (2014).}\label{fig3}
\end{figure}
Hotta (2014) has come up with a brilliant new idea for global computations of the convection zone, namely, to replace the usual anelastic approximation by a {\it reduced sound-speed} technique, in which the continuity equation is written as
\begin{equation}
\frac{\p \rho}{\p t}=-\frac{1}{\xi(r)^2}\div (\rho_0\bv).
\end{equation}
An example of one of his numerical experiments is shown in Fig. \ref{fig3}.

We also heard interesting talks about oscillatory dynamos without an omega-effect from Masada-san (2013), the effect of turbulent pumping on the solar cycle from Dibyendu Nandy (2013), and stellar dynamos in which buoyant loops are generated by a spot dynamo from Sacha Brun (2013).

\section{The Photosphere and Chromosphere}

Hiroko Watanabe (2012) gave an interesting review of the properties of umbral dots. As the magnetic field increases, their size and rise speed decreases but their lifetime remains the same.  They tend to cluster at the edges of the strongest umbral field, and it is possible in future that comparison with models may indicate properties of the subsurface field.

\begin{figure}
 \begin{center}
  \includegraphics[width=8cm]{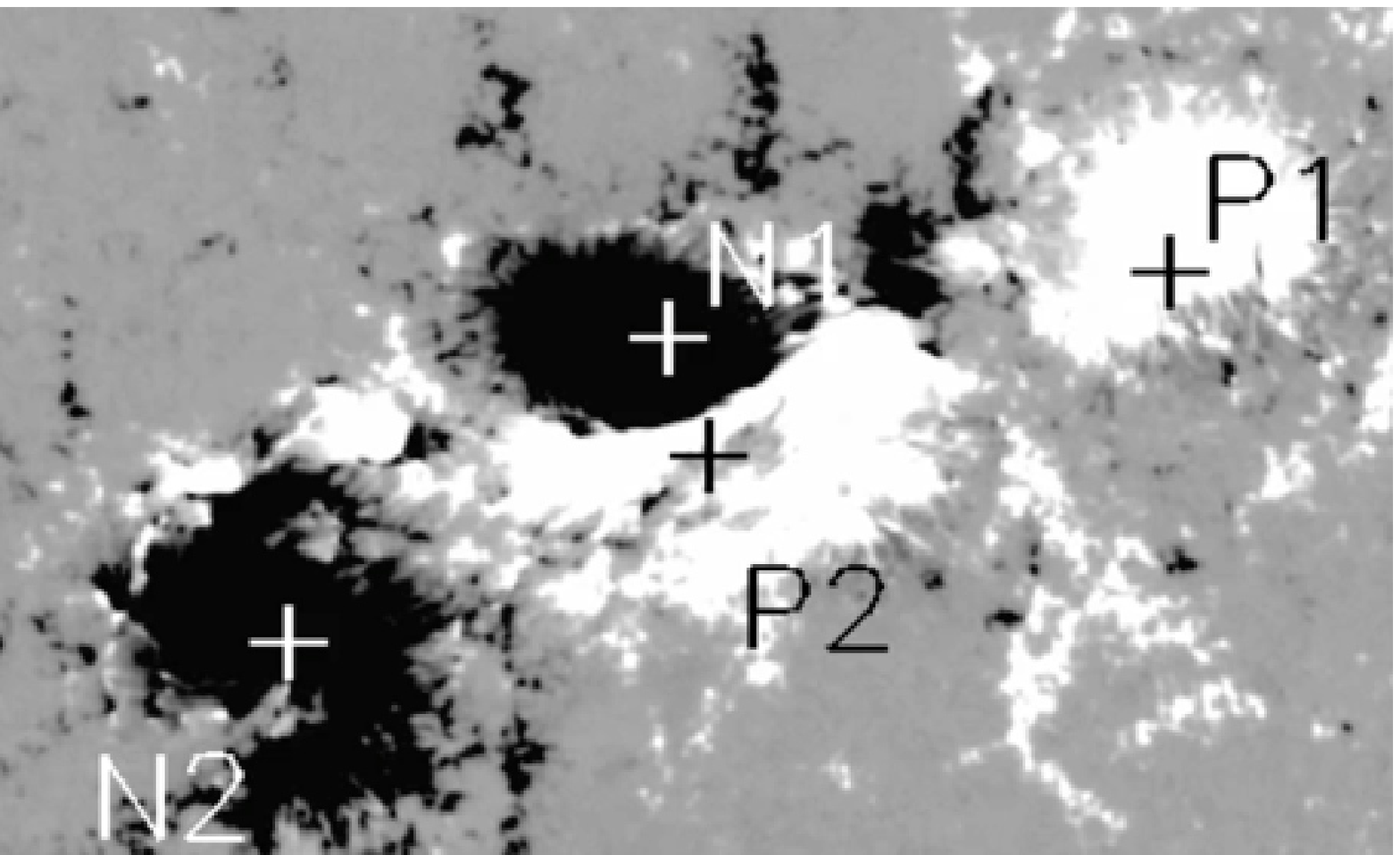}
  \includegraphics[width=8cm]{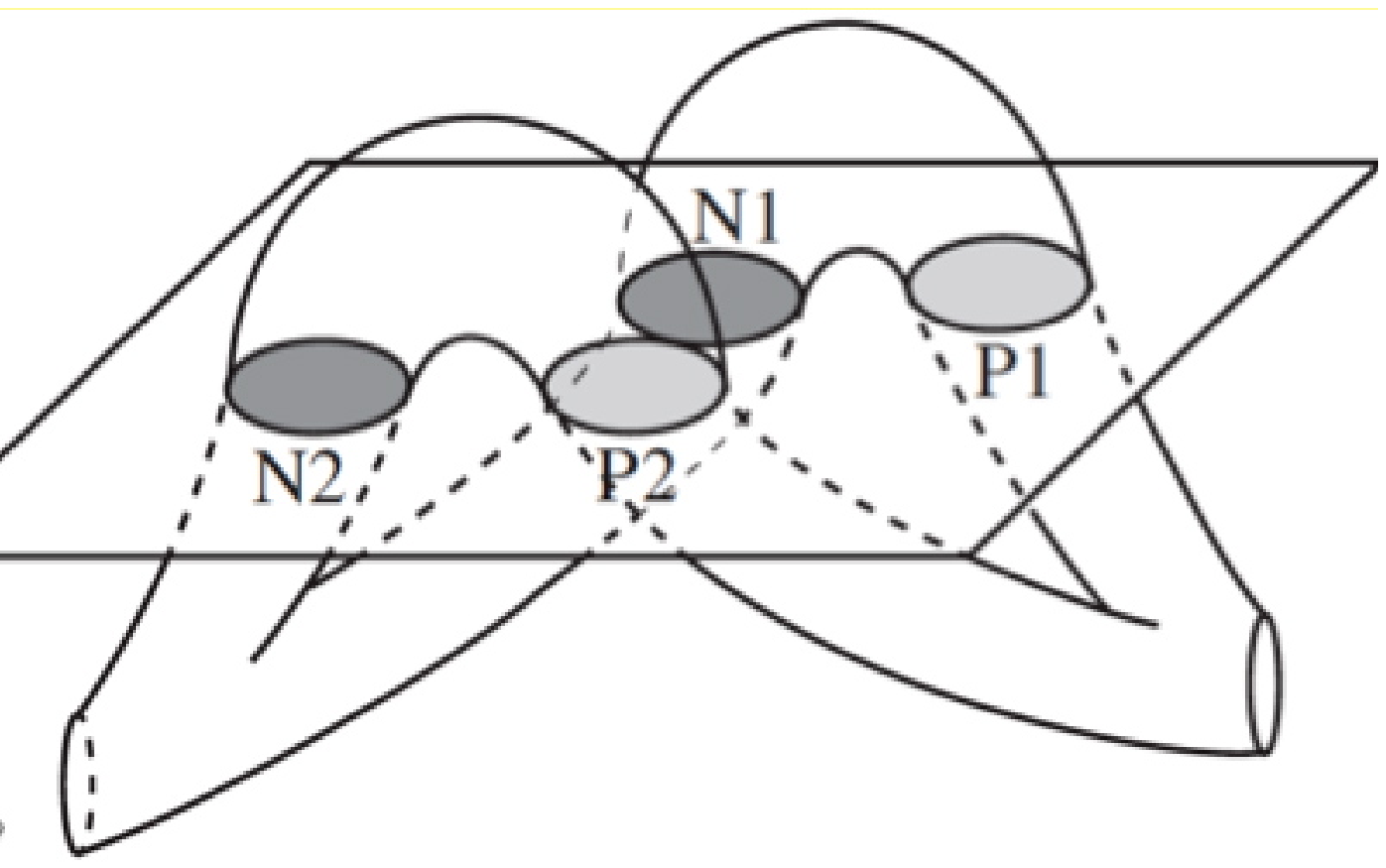}
 \end{center}
\caption{(a) Magnetogram of a quadrupolar active region in which the central two spots are sliding past one another. (b) A cartoon of the inferred split flux tube structure.}\label{fig4}
\end{figure}
Shin Toriumi, another rising star, reviewed flux emergence (Toriumi 2013, 2014).  He first described observations and simulations of emergence from the deep interior, and then suggested that resistive processes are important in the birth of active regions. Finally, he gave a recent example  of the formation of a flaring active region in which he inferred from sunspot motions that there may well have been a single flux tube below the photosphere which split into two parts (Fig.\ref{fig4}).

In future, since simulations have shown the difficulty of encouraging flux to emerge completely through the photosphere,  it would be good to try and estimate from observations and theory just how much flux is likely to pile up below the photosphere, both in the quiet Sun and in active regions. The presence of such flux may affect and interact with the near-surface shear layer, and it may also provide a background seed on which convection can operate  and generate granular magnetic loops.

\begin{figure}
 \begin{center}
  \includegraphics[width=8cm]{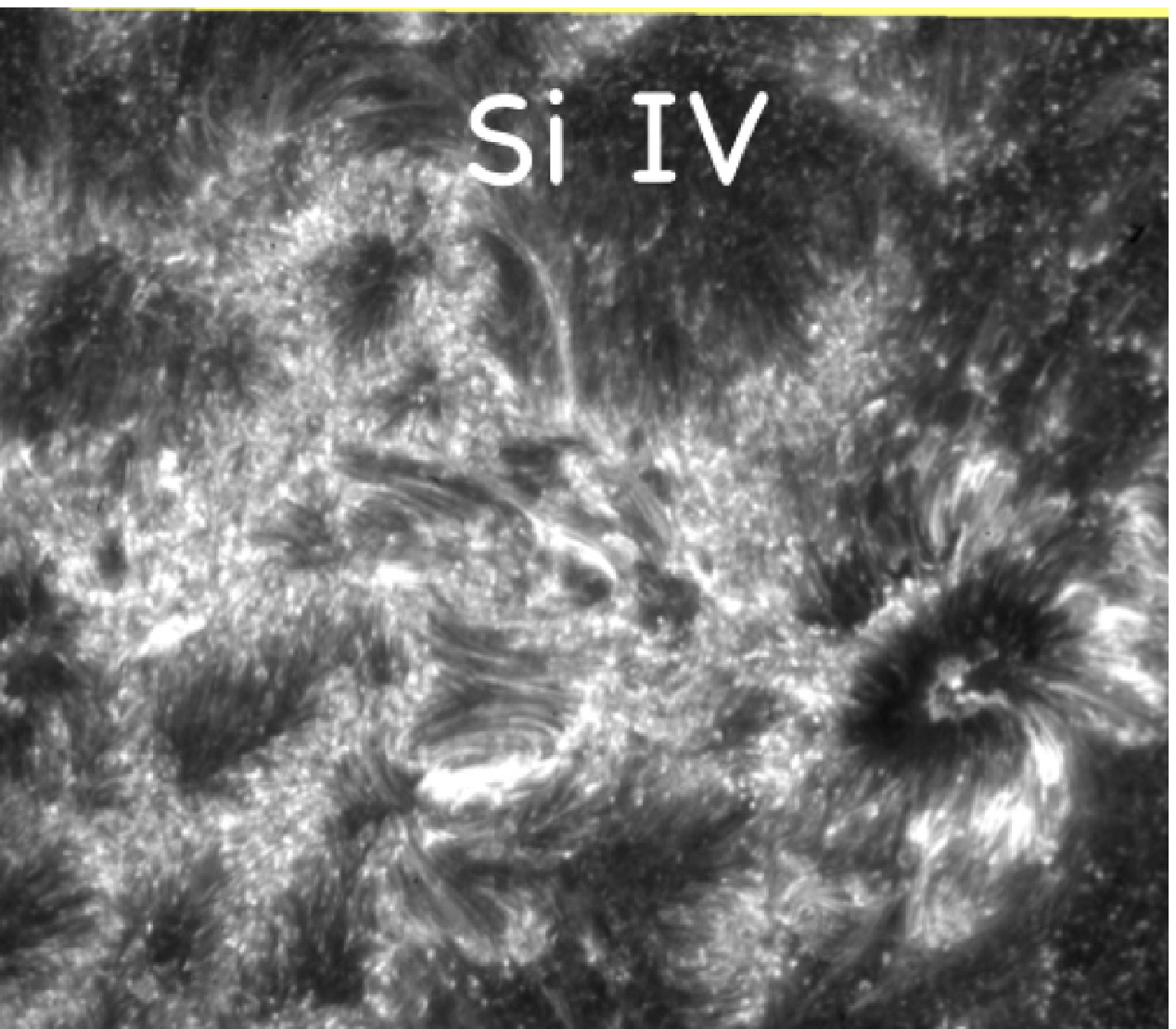}
  \includegraphics[width=7.3cm]{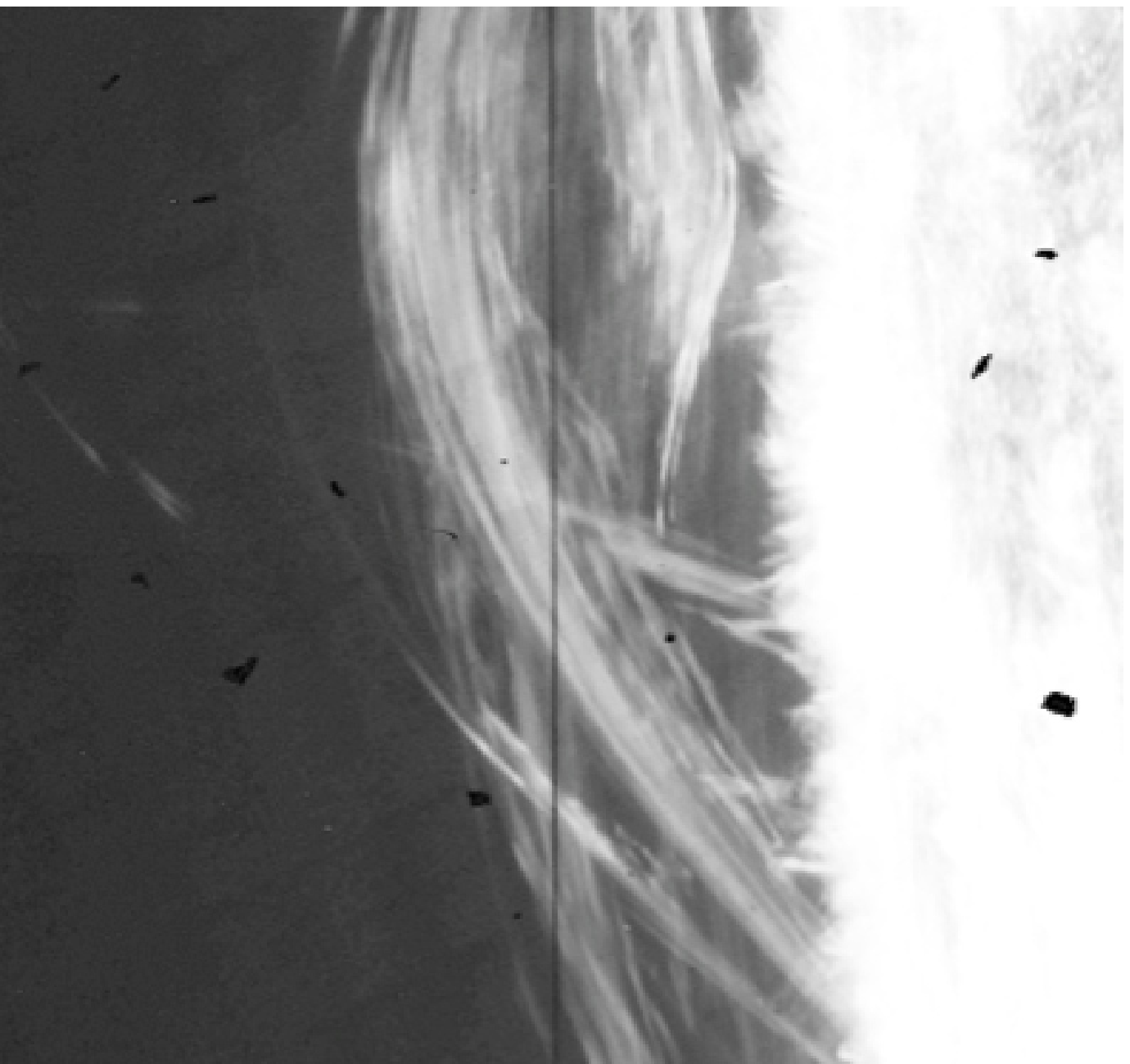}
 \end{center}
\caption{Images of fine-scale structuring of the solar atmosphere from IRIS (a) from above and (b) from the side.}\label{fig5}
\end{figure}
Alan Title (2013) presented some stunning UV slit jaw images and spectra from the new mission IRIS (Fig.\ref{fig5}).  These show that the chromosphere is incredibly dynamic, with rapid fine-scale brightnings and motions everywhere.  Ted Tarbell (2013) described coordinated observations from IRIS, SST and Hinode, while Tiago Pereira (2013) focussed on spicules, Viggo Hansteen compared the presence of a multitude of cool loops with models, and Mark Cheung showed examples of recurrent helical jets.   Clearly in future it is important to try and determine the causes of such fine-scale dynamics.

We also heard a variety of other talks about the photosphere. For example, Luis Bellot Rubio (2012) described the latest results about the ubiquitous horizontal magnetic fields discovered with Hinode with typical fields of 140 Gauss.  Daiko Shiota (2012)  presented details of the reversal of polar magnetic fields, while Ada Ortiz Carbonell (2014) described an example of granular flux emergence in the form of a magnetic bubble.  Andreas Lagg showed us the properties of granules in a light bridge, while Yukio Katsukawa (2012) presented details of photospheric power spectra.  Finally, two more future stars to watch out for in our field, Sanja Danilovic (2013) and David Buehler (2013), discussed the complex properties of 2D magnetic inversions for internetwork Hinode/SP data and for plage flux tubes.

\begin{figure}
 \begin{center}
  \includegraphics[width=10cm]{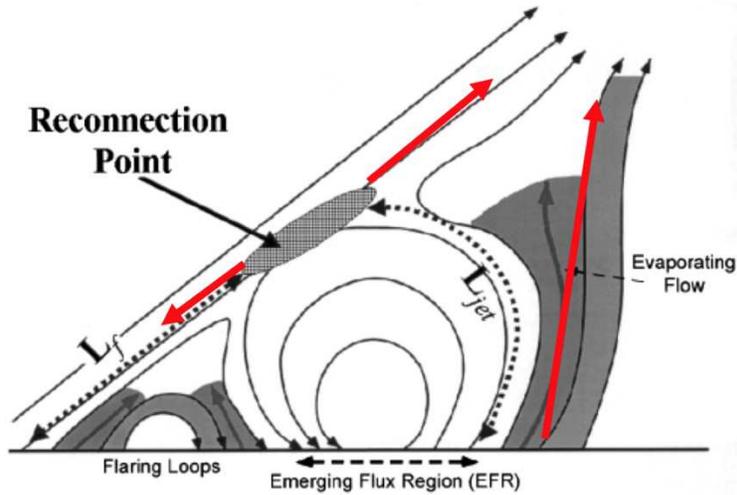}
 \end{center}
\caption{A numerical experiment by Shimojo and Shibata (2000) on jets produced by 2D reconnection.}\label{fig6}
\end{figure}
\section{Coronal Structure and Heating}

Etienne Pariat (2012) gave a masterly review of coronal jets, splitting them into {\it standard jets} and {\it blowout jets}, which are more complex, arise from multipolar magnetic fields and often have a cool part.  Often (anemone) jets occur at 3D null points by spine-fan reconnection, and helical structure is common.  He described the basic 2D mechanism first suggested by Heyvaerts, Priest and Rust (1977) and subsequently developed by Forbes and Priest (1984) as well as Shibata (1992), Yokoyama (1996) and their colleagues (Fig.\ref{fig6}).  These suggested a hot fast jet produced by reconnection together with a cooler jet produced by evaporation.

\begin{figure}
 \begin{center}
  \includegraphics[width=15cm]{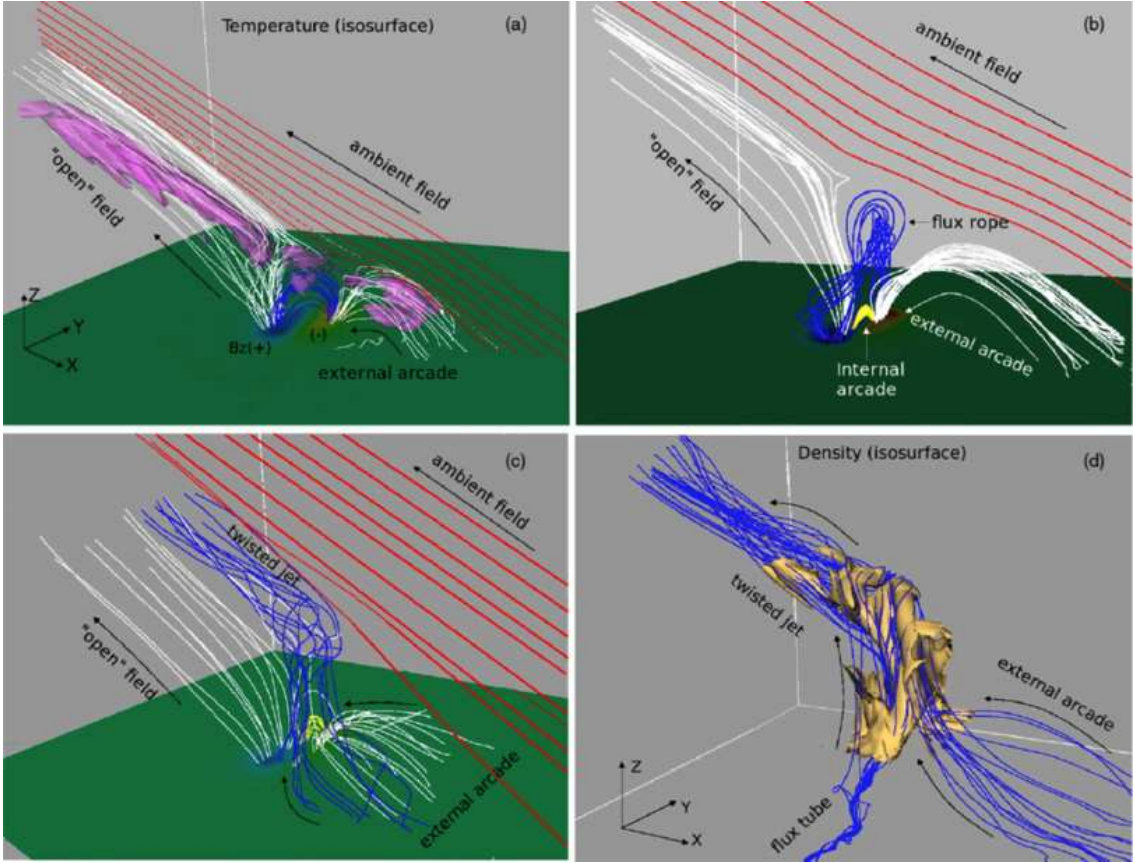}
 \end{center}
\caption{A numerical experiment by Archontis (2013a, 2013b) on jets produced by 3D reconnection, showing (a) the temperature, (b) the magnetic fields, (c) a twisted jet and (d) the density.}\label{fig7}
\end{figure}
In three dimensions, the process is more complex and has new features, such as untwisting jets, according to experiments by Pariat (2010) and Archontis (2013a,b), and it is still uncertain whether the cooler chromospheric jets are produced by slow-mode shocks or pressure build-up. 

In future, we need more observations and numerical experiments on the 3D aspects of jets produced by reconnection, which could shed light on several fronts.  For example, what is the role of time-dependent jets in generating waves? What is the role of magnetic helicity? How much of the twist is releasing stored up twist and how much is due to the conversion of mutual magnetic helicity into self helicity by the reconnection process?  Furthermore, what is the effect of the jets on both coronal heating (converting kinetic energy into heat and spreading out the energy of hot jets) and on accelerating the solar wind?

Harry Warren (2013) gave an innovative talk about active-region coronal heating, stressing a promising technique (sparse Bayesian inference) to balance uncertainty and complexity in models.  He also showed a comparison of an observed SDO active region with a nonlinear force-free model, in which the temperature and density on 1000 field lines were calculated.  These suggest that the heating ($H$) has the following scaling
\begin{equation}
H \sim \frac{B}{L}
\end{equation}
with magnetic field ($B$) and loop length ($L$).  They also imply that the heating events occur on time-scales more rapid than the cooling time.

In future, there is room for much more comparison between theory and observation in order to determine the likely heating mechanisms at work.

\begin{figure}
 \begin{center}
  \includegraphics[width=13cm]{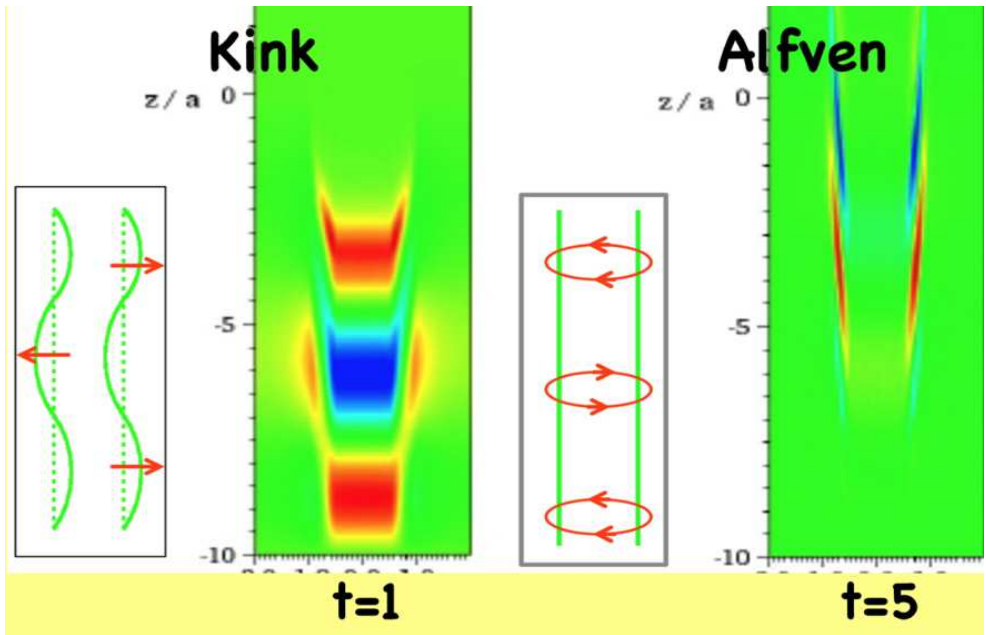}
 \end{center}
\caption{A numerical experiment by De Moortel and colleagues demonstrating the conversion as they propagate of kink waves into Alfv\'e n waves localised in a flux tube boundary.}\label{fig8}
\end{figure}

Ineke De Moortel reviewed wave heating of the corona (see De Moortel, 2012; Parnell and De Moortel, 2012), first of all describing observations of {\it Alfv\'enic waves} in spicules and in the corona (with COMP and SDO), which imply that 100 W m$^{-2}$ is required in the quiet Sun and 2000 W m$^{-2}$ in active regions.  She then pointed out that they could be generated  directly from photospheric vortices or by mode coupling from kink modes, in which the waves become localised in a flux tube boundary (Fig.\ref{fig8}). Their observational signature is at present unclear, since they could produce an impulsive or turbulent emission.

Waves are likely to be important in heating part of the corona, but there is a need to develop the basic theory beyond the current paradigm of simple flux tubes and to deduce observational signatures in more realistic geometry.

We also heard excellent talks on a variety of other topics. Marc de Rosa described the effect of spatial resolution on nonlinear force-free extrapolation.
On prominences, two more future stars are David Orozco Suarez (2014), who discussed the inferred magnetic field of prominence threads, and Andrew Hillier (2014), who showed how observations of rising plumes can be used to infer the plasma beta in a prominence.  In addition, Elena Dzifcakova (2014) gave interesting insights on the nature of the prominence transition region.

Regarding jets and flux emergence, Irina Kitiashvili (2013) described ejection by a photospheric vortex tube, while Vasyl Yurchyshyn (2013) suggested that spicules are accelerated by reconnection (Uchida-san would be pleased), and Len Culhane (2012) talked about the properties of solar wind outflow from active regions using EIS.  Shinsuke Takasao (2013) discussed jets accelerated by reconnection and shocks, and Peter Young (2014) introduced the idea of ``dark jets" in coronal holes. 

Finally, there were several talks on coronal heating in general.  Three other future stars were: Philippe Bourdin (2013), who presented a 3D model of an active region being heated in response to photospheric motions; Jiansen He, who talked about slow-mode waves and outflows from reconnection; and Hwanhee Lee, who discussed a variety of different magnetic configurations.  In addition, observations from the Hi-C rocket flight were shown by Sabrina Savage  of active-region dynamics  and by Paola Testa (2013) of moss variations due to nanoflares.

\section{Flares and CME's}

\begin{figure}
 \begin{center}
  \includegraphics[width=11cm]{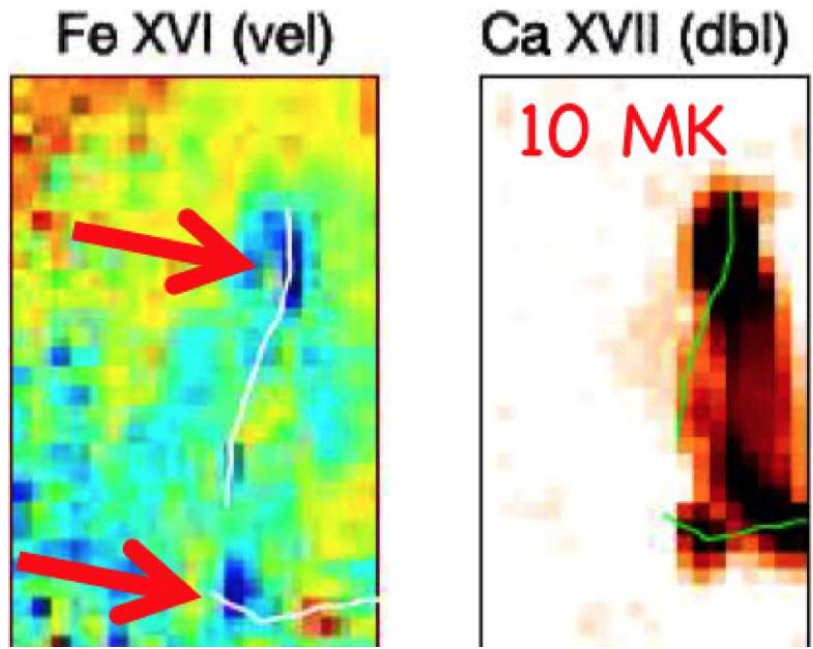}
 \end{center}
\caption{Observations by Del Zanna (2014) of a coronal loop with EIS, showing (a) Doppler shifts and (b) 10 MK plasma.}\label{fig9}
\end{figure}
Helen Mason gave a masterly review of evaporation in small flares using imaging and spectroscopy from Hinode/EIS.  In the 90's, Doschek had observed blue shifts during flares with BCS, but at the time he had no idea about their location.  Now with EIS, Del Zanna and Mason (2013, 2014) have shown how EIS blue shifts are located in kernels at the ends of hot coronal loops (Fig. \ref{fig9}). They occur only in lines at 2--3 MK and represent evaporation from the chromosphere. The upflowing plasma is located at a height of 200 km with a density of 10$^{11}$ cm$^{-3}$ and its properties agree with those from a conduction-driven 1D simulation.   SDO/EVE has produced more examples of upflows at 100--200 km s$^{-1}$.

In future, it would be interesting to study such evaporation processes in large more-complex flares and to try and determine the heating and particle acceleration mechanisms.

\begin{figure}
 \begin{center}
  \includegraphics[width=7cm]{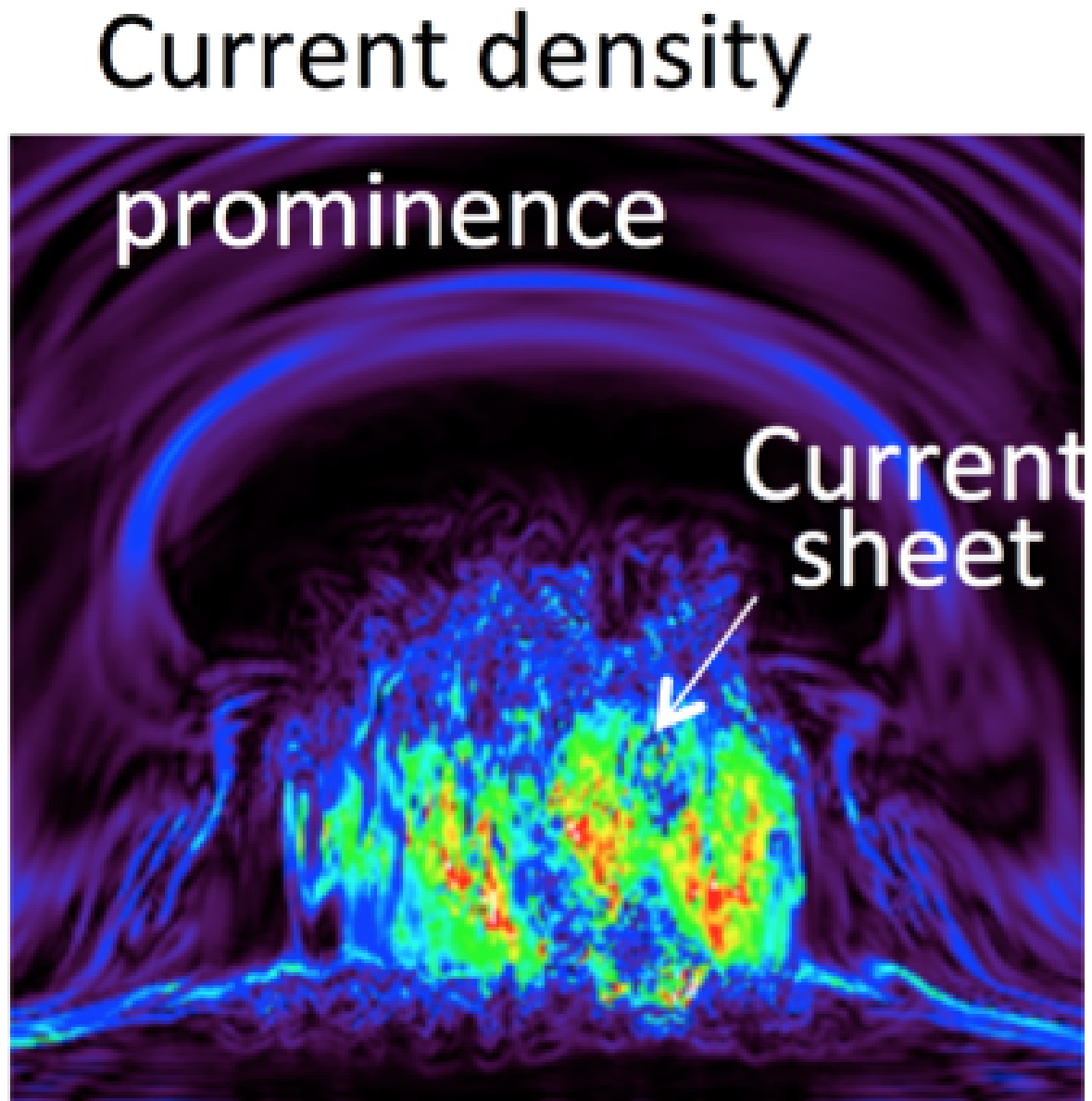}
    \includegraphics[width=9cm]{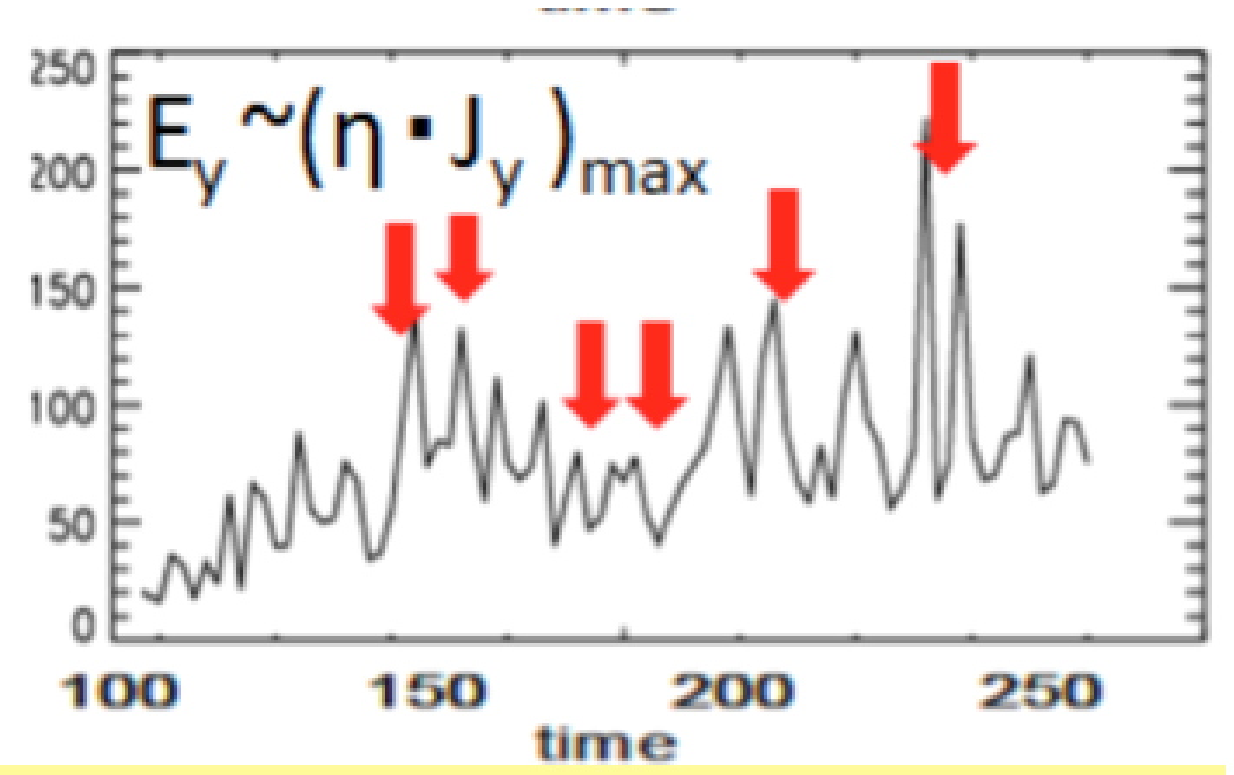}
 \end{center}
\caption{3D simulations by Nishida et al (2013) of a current sheet below and erupting prominence, showing (a) a snapshot of the current density and (b) the electric field as a function of time}\label{fig10}
\end{figure}
The subject of reconnection and particle acceleration in eruptive flares was reviewed by Naoto Nishizuka (2013) (also a rising star). He showed how brightnings start below an erupting prominence, and suggested that the eruption drives impulsive reconnection in a current sheet.  The fragmentation and ejection of plasmoids in the sheet has been demonstrated in 3D simulations (Fig.\ref{fig10}), and test-particle orbits have found Fermi acceleration at a fast-mode shock as well as stochastic acceleration at multiple separators.

In future, it would be good to compare with observations of SAD's (supra-arcade downflows) and also to develop self-consistent plasma physics of the process.

\begin{figure}
 \begin{center}
  \includegraphics[width=5cm]{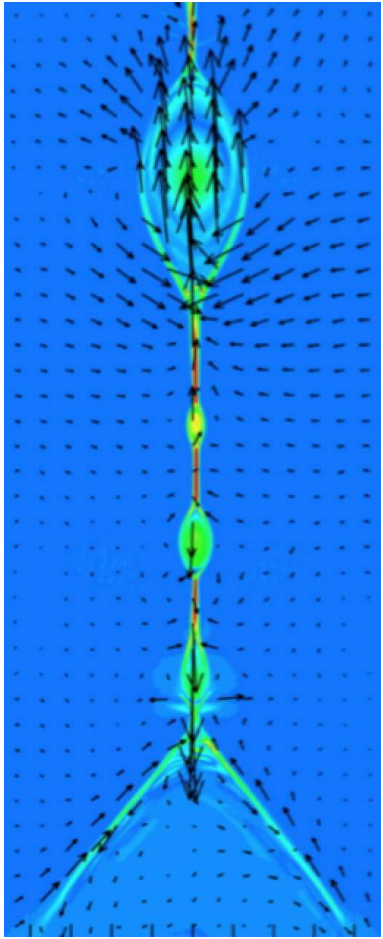}
 \end{center}
\caption{Fragmentation of the current sheet below a CME to give a series of plasmods, as shown in a current density plot from a simulation by Mei (2012).}\label{fig11}
\end{figure}
Then Jun Lin discussed large-scale current sheets in CME's, showing how they have been observed in LASCO images, with plasmoids and typical thicknesses of 10$^3$ km, lengths of  10$^5$ km and Alfv\'en Mach numbers of 0.01.  Numerical experiments reveal fragmentation and the properties of plasmoids with upflow and downflow velocities of 150--250 km s$^{-1}$ and 90--160 km s$^{-1}$, respectively.  At a magnetic Reynolds number of 10$^5$, there are typically 15 plasmoids and the turbulence enhances reconnection.

In future, determining the nature and properties of the turbulence will be helpful, as well as determining the formation mechanism for the the blobs (such as perhaps secondary tearing).

Other interesting talks that we heard were about: the location of non-thermal velocities from EIS (by Louise Harra, 2013); an MHD eruption (by Ed De Luca, 2013); flare ribbons and current ribbons by (Miho Janvier, 2013, 2014 , another rising star); shear flow with SOT along a polarity inversion line (by Toshi Shimizu, 2013); a flare observed with FISS/NST (by Hyungmin Park, 2013); the way in which SAD's indicate fragmentation of a turbulent flare current sheet (by David McKenzie, 2013, and Kathy Reeves, 2013); reconnection outflows in an X-flare (by Hirohisa Hara, 2011); supersonic outflows in a prominence eruption (by David Williams, 2013); the magnetic field deduced from EIT waves (by David Long, 2013); and hard X-rays from FOXSI by Shin Ishikawa (see Krucker et al, 2011).

\section{Superflares on Solar-Type Stars}

Karel Schrijver (2012) and Hiroyuki Maehara (2012) reviewed the concept of superflares. On the Sun, the flare energy is mostly in white light and so is hard to measure. From Kepler, the frequency of stellar flares fits a power law and  increases with rotation.  The flare energy is independent of rotation but depends on spot area and so a superflare needs superspots. On the Sun, one flare at 10$^{34}$ erg is expected every 800 years and one at  10$^{35}$ erg every 5000 years, so space weather can become much worse.

We also heard how Hinode helps understand stellar flares (from Petr Heinzel), details of stellar winds from young solar-type stars (from Takeru Suzuki, 2013), how a magnetic storm twice as large as the Carrington event is possible (from Bruce Tsurutani, 2014) and about rapid events in tree rings (from Fusa Miyake, 2013).

\section{In Future}

\begin{figure}
 \begin{center}
  \includegraphics[width=7cm]{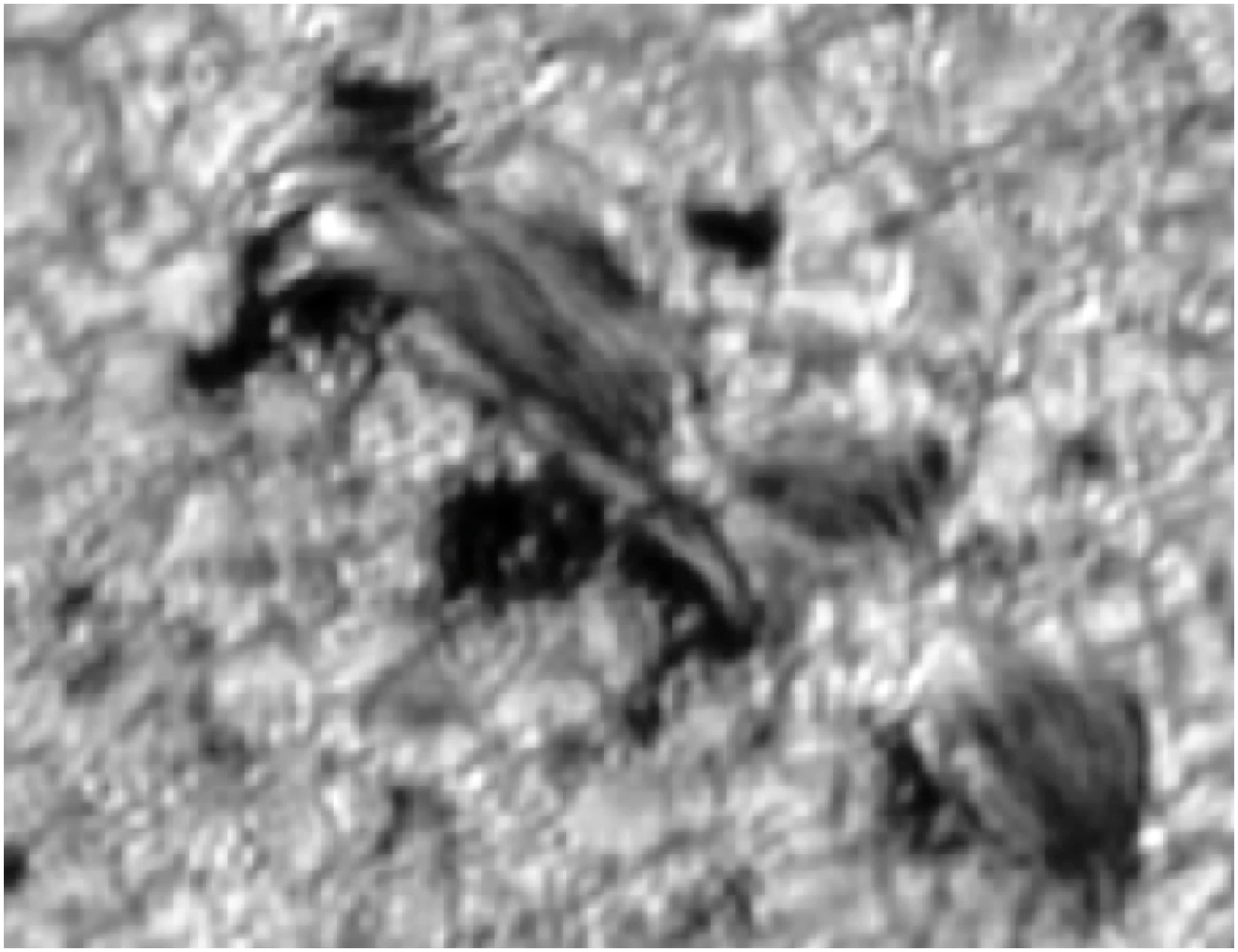}
    \includegraphics[width=9cm]{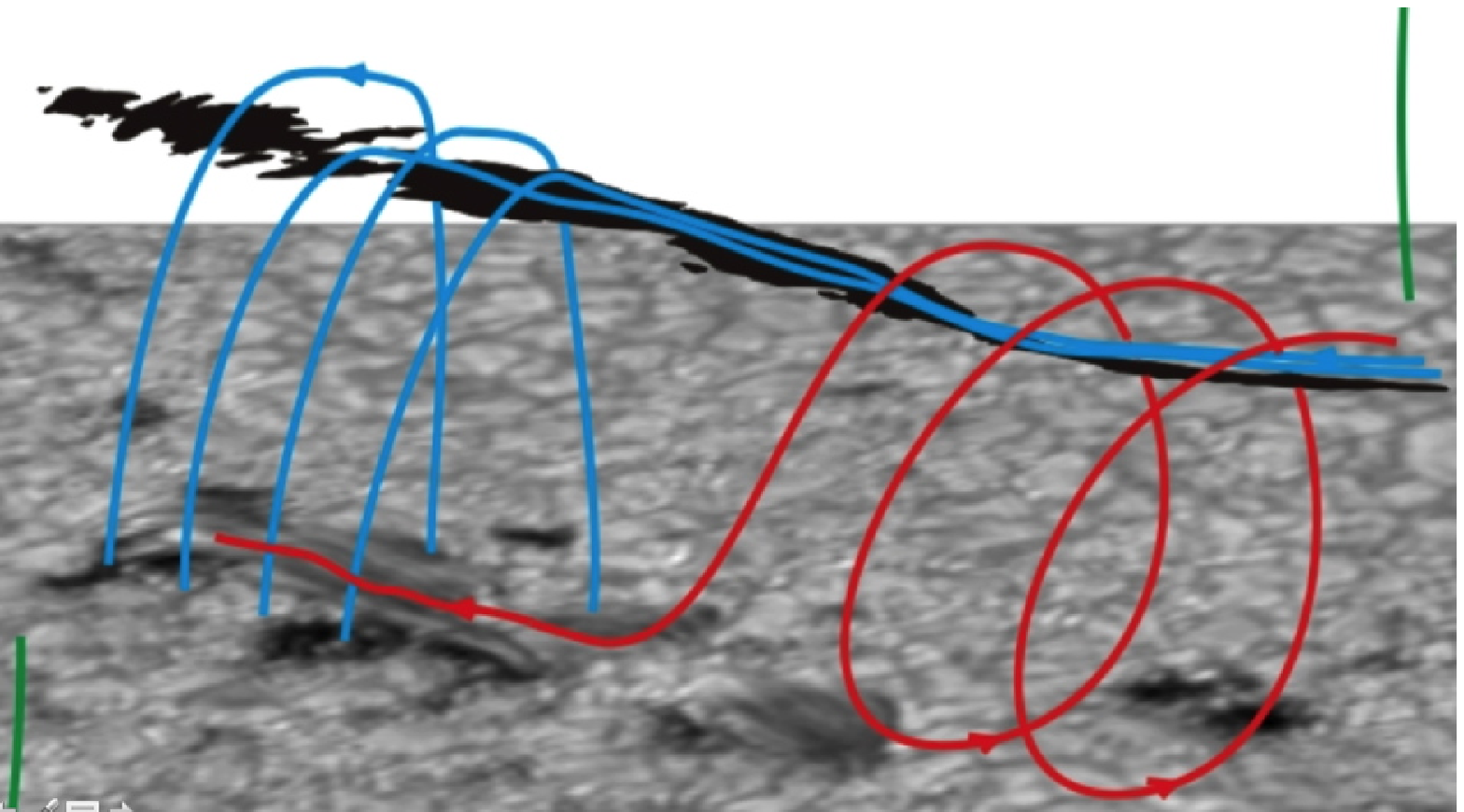}
 \end{center}
\caption{(a) An example of orphan penumbrae and (b) a suggested magnetic configuration}\label{fig12}
\end{figure}
Brian Welsh (2014) raised three questions about the nature of photospheric magnetic fields and flows. First of all, do photospheric flows drive Alfv\'enic turbulence to heat the atmosphere?  He does find flows that are faster and shorter-lived at smaller scales, but the observed flows do not agree with the Van Ballegooijen (2014) model.  Secondly, is flux emergence ideal or is reconnection necessary?  He often finds that flux loss, cancellations and the deduced electric fields suggest  reconnection is indeed important.   Thirdly, what is the cause of the changes in photospheric magnetic field during flares?  Perhaps a change in magnetic tension or a relation to sunquakes is involved.

In order to solve these questions properly, higher-resolution observations from Solar C and ATST are needed, together with new ideas and computational experiments.

Valentin Martinez Pillet (2014) raised the question: do you need continuous magnetogram observations for days with very high resolution and a small field of view?   He suggested that the answer is yes, but that you also need full-disc observations to provide the context.  He discussed an example of the puzzling nature of orphan penumbrae when sunspots have an unusual appearance (Fig. \ref{fig12}a).  He suggested that large field of view observations enable an understanding of such localised behaviour.  Thus, orphan penumbra occur when a strong horizontal magnetic field in the photosphere inhibits convection.  This may occur either during the emergence of active regions or when the axis of a twisted flux rope dips down to the photoshere (Fig. \ref{fig12}b).

We also heard about coronal loop strands from EIS, AIA and Hi-C (from David Brooks, 2013), about the plans for CLASP (from Ryohko Ishikawa, 2014), and about NST observations (from Sasha Kosovichev, 2013).

\section{Conclusion}

\begin{figure}
 \begin{center}
  \includegraphics[width=13cm]{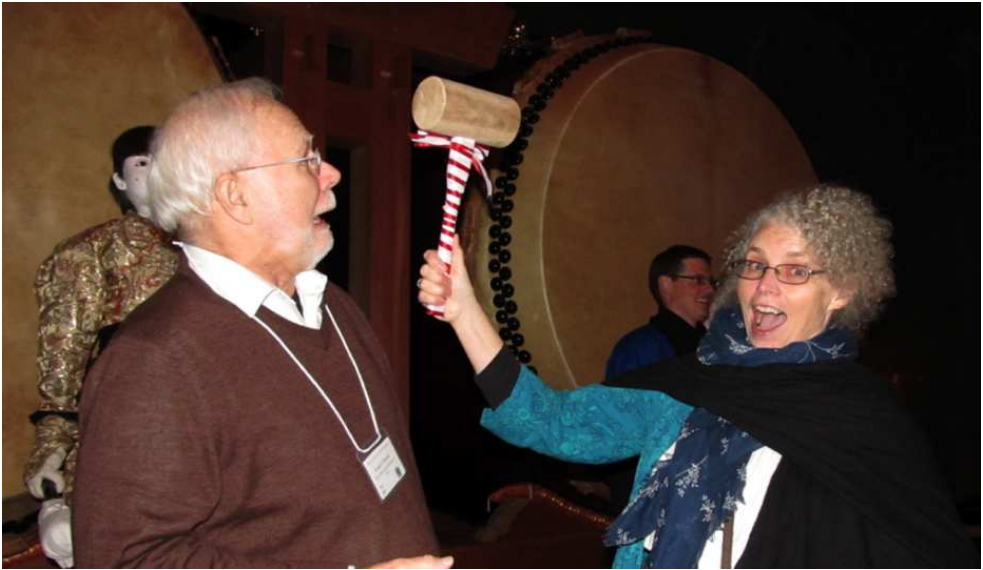}
 \end{center}
\caption{Helen beating sense into George at the banquet.}\label{fig13}
\end{figure}
Several amusing or memorable comments were made during the conference.  George Doschek asked ``Should we think out of the box or drink out of the box?"  But thankfully Helen beat sense into him (Fig. \ref{fig13}).  Shibata-san impressed us with dramatic Kitaro music accompanying flares. Watanabe-san told us about Kyoto's Galileo, called Mr Suzuki.  Ada Ortiz Carbonell talked about ``the most awaited romance".  Andreas Lagg daringly put the word ``naked" into his talk title.  Andrew Hillier had his father as a co-author. Louise Harra admitted to stopping us from going for our beer. Miho Janvier said ``If you don't follow my talk, blame the conference dinner alcohol".  

\begin{figure}
 \begin{center}
  \includegraphics[width=10cm]{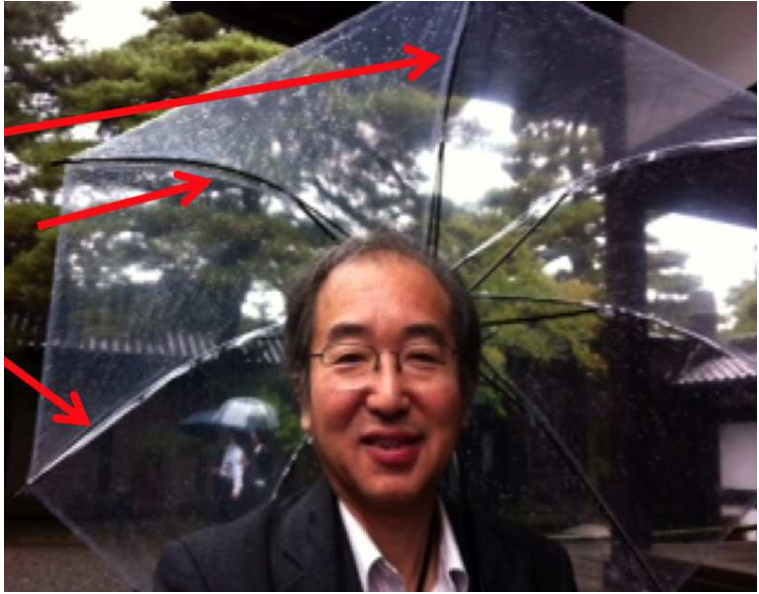}
 \end{center}
\caption{Shibata-san with a halo that possesses separatrices.}\label{fig14}
\end{figure}
\begin{figure}
 \begin{center}
  \includegraphics[width=15cm]{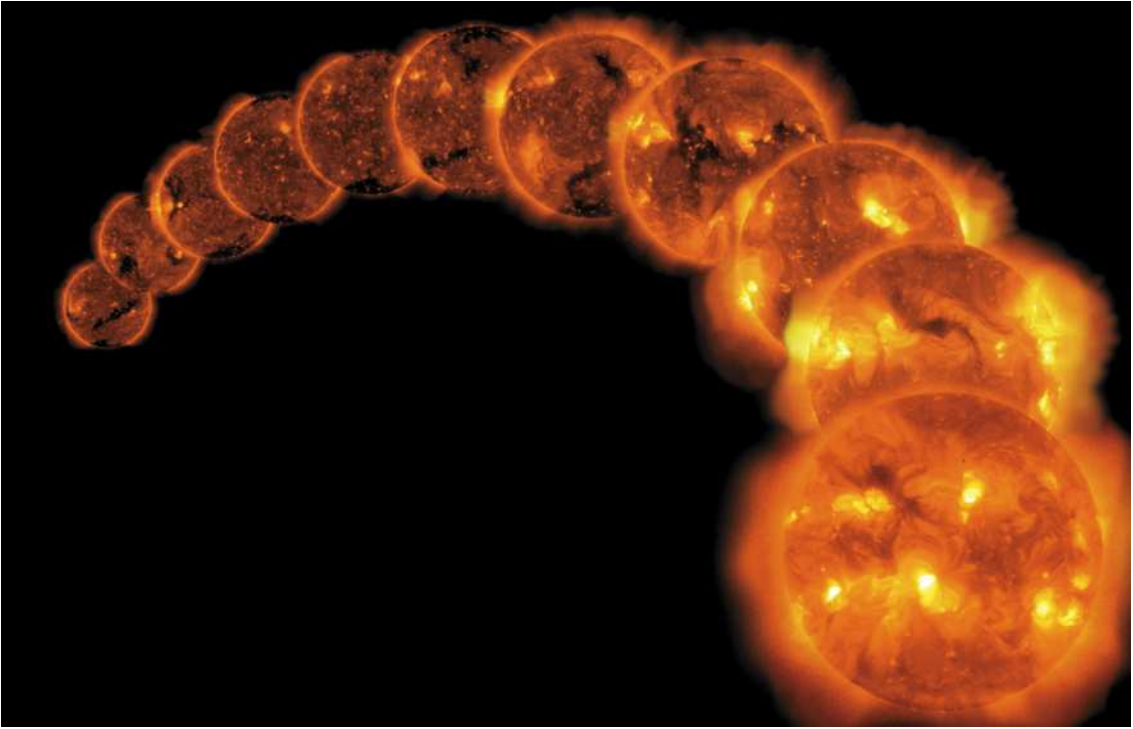}
 \end{center}
\caption{The corona from Hinode (courtesy Narukage and Tsuneta).}\label{fig15}
\end{figure}
We all agreed this had been a memorable conference in a beautiful location and  were most grateful to the members of the SOC for their hard work, led by Shibata-san (Fig. \ref{fig14}), and also to the LOC led by Ichimoto-san and especially to Shin'ichi Nagata.  But, as we parted we remembered to enjoy the beauty of the solar corona (Fig. \ref{fig15}).


\bigskip

\section{Acknowledgement} 
I am extremely grateful to Hirohisa Hara and Kazunari Shibata for hosting my visits to Tokyo and Kyoto, respectively, and for looking after me so well.  It was a real delight to meet old friends and make new ones.  I am also grateful to the University of Tokyo and to the Leverhulme Trust for financial support.



\end{document}